\documentclass[reprint,aps,prl,superscriptaddress,floatfix]{revtex4-2}

\usepackage[utf8]{inputenc}
\usepackage[T1]{fontenc}
\usepackage{graphicx}
\usepackage{amsmath,amssymb,bm}
\usepackage[dvipsnames]{xcolor}
\usepackage{hyperref}
\hypersetup{
	colorlinks=true,
	linkcolor=black,
	citecolor=black,
	urlcolor=black,
	pdfborder={0 0 0},
	breaklinks=true
}
\usepackage{physics}

\begin{document}

\title{Dissipative dynamics and superradiant continuous time crystal in a \textcolor{black}{Rydberg-interacting} Dicke system}

\author{Haohang Zhou}
\affiliation{State Key Laboratory of Photonics and Communications, School of Physics and Astronomy, Shanghai Jiao Tong University, Shanghai 200240, China}

\author{Xianfeng Chen}
\affiliation{State Key Laboratory of Photonics and Communications, School of Physics and Astronomy, Shanghai Jiao Tong University, Shanghai 200240, China}
\affiliation{Collaborative Innovation Center of Light Manipulations and Applications, Shandong Normal University, Jinan, 250358, China}

\author{Luqi Yuan}
\email{yuanluqi@sjtu.edu.cn}
\affiliation{State Key Laboratory of Photonics and Communications, School of Physics and Astronomy, Shanghai Jiao Tong University, Shanghai 200240, China}

\begin{abstract}
	
The interplay between many-body interactions and controlled dissipation provides a rich setting for nonequilibrium quantum phases.
We explore the superradiant phase transitions in a Rydberg-interacting open Dicke model with cavity loss and \textcolor{black}{local dissipation}. We find that
\textcolor{black}{the nonlinear feedback from atomic interactions stabilizes the atomic collective excited state and induces resonant collective motion, which can be further converted into robust superradiant oscillations with the inclusion of local dissipation.}
\textcolor{black}{Beyond the mean-field approximation, we prove this system can support the finite-size local dissipative temporal order, i.e., continuous time crystal, from evidences of the decreasing damping of a finite--frequency Liouvillian mode and the linear-size scaling of quantum fluctuations.}
\textcolor{black}{The correlation between the atomic colletive dynamics and the cavity photon number makes our findings easy to detect from output optical signal.}
By bridging interacting atomic matter and driven-dissipative cavity QED, our 
\textcolor{black}{Rydberg-interacting Dicke model provides a unique spin-$1/2$ platform for exploring dynamical phase transitions and macroscopic temporal order in open quantum matter.}
\end{abstract}

\maketitle

\paragraph{Introduction.}
The open Dicke model is a paradigmatic system for studying collective quantum light-matter physics, capturing phenomena ranging from superradiance and bistability to nonequilibrium dynamical phase transitions \cite{PhysRevA.75.013804,PhysRevA.85.013817}. Upon its cavity-QED realizations \cite{Clerk2010RMP}, photon leakage in the cavity can lead to the intrinsic driven-dissipative dynamics, resulting in nonequilibrium steady states that can be directly read out via input--output measurement in experiments \cite{GardinerCollett1985,Dreon2022Nature, PhysRevLett.133.106901}. Therefore, considerable progresses have been achieved along this line, including experimental implementations with ultracold atoms in optical cavities \cite{nature09009,PhysRevLett.104.130401,PhysRevLett.127.253601} and extensive theoretical studies of nonequilibrium dynamical phase transitions, bistability and hysteresis, and collective oscillations \cite{RevModPhys.85.553,PhysRevA.102.063702,PhysRevLett.131.113602,mz92-6l9g,42rz-xhxz}.All these advances have established the open Dicke framework as a broad platform for exploring nonequilibrium collective physics, quantum simulation and dissipative state engineering \cite{Mivehvar02012021}.

The standard open Dicke model is governed primarily by photon-mediated collective couplings. \textcolor{black}{Direct many-body interactions, however, are the defining ingredient of collective quantum matter, as they drive quantum phase transitions and fundamentally alter nonlinear dynamics far from equilibrium} \cite{RevModPhys.95.035002,Polkovnikov2011RMP,Eisert2015NatPhys}. \textcolor{black}{A range of experimental tools now exist for introducing direct interactions into cavity-QED, from Rydberg dressing to dipole--dipole coupling schemes} \cite{Weckesser2025Science,Zeiher2016NatPhys,PhysRevA.82.033412,PhysRevLett.131.063401}. Therefore, a fundamental question arises naturally: what is the impact on the driven-dissipative phenomena when \textcolor{black}{Rydberg interactions} are introduced into the open Dicke physics? 

\textcolor{black}{In this Letter, we investigate how Rydberg interactions reshape the dynamics of an open Dicke model subject to both cavity loss and local dissipation, which generates a rich phase structure encompassing stationary, multistable, and oscillatory regimes.} Among these phases, a superradiant continuous time crystal (CTC) emerges from the interplay between dissipation and interaction-induced nonlinear feedback in our system, distinct from earlier discrete time crystals that rely on periodic driving and subharmonic response~\cite{PhysRevLett.120.040404,Kessler2020NJP}. \textcolor{black}{At the mean-field level, Rydberg interactions introduce a second critical coupling, i.e., the collective excited state survives as a steady state, its instability threshold varying with the collective interaction, producing the bistable window. When the thresholds of the ground state and the excited state turn to be same, an oscillation phase occurs. Importantly,  we find that, in our model, the local atomic dissipation can stabilize a robust CTC phase, while in general Dicke system, local dissipations usually suppress temporal order ~\cite{RussoPohl2025PRL,jhd4-1khw}.} 
\textcolor{black}{We provide further calculations beyond the mean-field approximation, by taking second-order cumulant-expansion (SCE) and wave-function Monte Carlo (WFMC) methods, to verify the dissipation-induced collective temporal order characteristics still exist under high-order quantum effects and finite-size system }~\cite{Iemini2018PRL,jhd4-1khw}.
\textcolor{black}{Our findings in this Rydbery-interacting Dicke model can be directly reflected in the cavity output, where the photon number is correlated with the collective atomic dynamics, and thus provides an experimentally accessible optical readout.}

\begin{figure}[tp]
	\centering
	\includegraphics[width=0.98\linewidth]{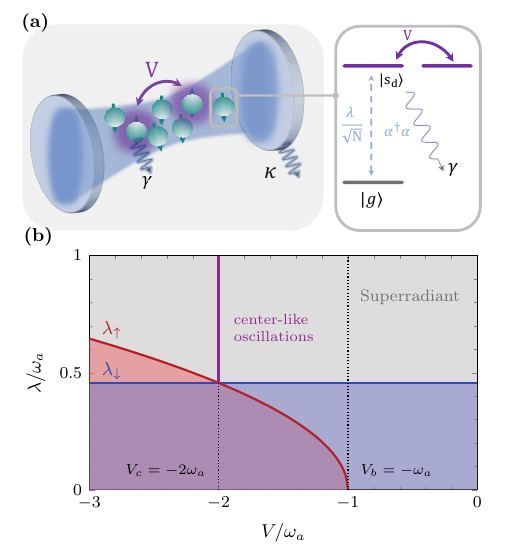}
	\caption{\textcolor{black}{Effective Rydberg-interacting open Dicke model and its mean-field phase structure without local dissipation.
	(a) Effective two-level atoms interact with strength $V$ and couple collectively to a cavity with photon loss $\kappa$ and atomic local dissipation $\gamma$.
	(b) Mean-field phase diagram at $\gamma=0$.
	The black and red boundaries give the critical couplings $\lambda_{\downarrow}$ and $\lambda_{\uparrow}$; at $V_c=-2\omega_a$, they coincide and yield center-like oscillations above $\lambda_c$.}}
	\label{fig:setup}
	\label{fig:mf}
\end{figure}
\paragraph{Model.}
We consider an ensemble of $N$ identical \textcolor{black}{effective two-level} atoms coupled to a single-mode optical cavity, as sketched in Fig.~\ref{fig:setup}(a).
\textcolor{black}{We denote the two states by $|g\rangle$ and $|s_d\rangle$, and assume an effective Rydberg interaction between atoms occupying $|s_d\rangle$.}
\textcolor{black}{In the Supplemental Material, we outline a possible scheme based on Rydberg dressing~\cite{SM}.}
\textcolor{black}{The effective Rydberg-interacting dissipative Dicke model is described by the Hamiltonian (setting $\hbar=1$):}
\begin{equation}
H=\omega_c a^\dagger a
+\omega_a \sum_{i}^N n_i
+\frac{\lambda}{\sqrt{N}}\sum_{i}^N (a^\dagger+a)\sigma_i^x
+\sum_{i\neq j}\frac{U_{\rm eff}(r_{ij})}{2}n_i n_j .
\label{eq:H_eff_prl}
\end{equation}
Here $a$ is the cavity photon annihilation operator, $\sigma_i^x=|g_i\rangle\langle s_{d,i}|+\mathrm{h.c.}$ is the Pauli operator acting on the effective two-level subspace $\{|g\rangle,|s_d\rangle\}$, and $n_i\equiv |s_{d,i}\rangle\langle s_{d,i}|$ is the \textcolor{black}{upper-state} population operator. $\omega_c$ is the effective cavity \textcolor{black}{detuning}, and $\omega_a$ is the effective atomic splitting \textcolor{black}{between the two atomic levels}.
The collective coupling $\lambda$ \textcolor{black}{parameterizes the Dicke light-matter interaction and} is scaled as $\lambda/\sqrt{N}$ in Eq.~\eqref{eq:H_eff_prl}.
The last term denotes \textcolor{black}{an effective Rydberg interaction $U_{\rm eff}(r_{ij})$ between atoms occupying $|s_d\rangle$}~\cite{SM}.
In the rest of this work we employ a permutation-invariant all-to-all approximation,
$U_{\rm eff}(r_{ij})\approx V/(N-1)$, where the collective interaction $V\equiv\sum_{j(\neq i)}U_{\rm eff}(r_{ij})$ is a constant parameter \cite{jhd4-1khw,SM}.
We also introduce collective spin operators
$S_a\equiv \frac12\sum_{i=1}^N\sigma_i^a$ ($a=x,y,z$), such that $\sum_i\sigma_i^x=2S_x$ and $\sum_i n_i=\frac{N}{2}+S_z$.
\textcolor{black}{To complete the effective open-system description, we include cavity photon leakage and atomic local dissipation from $|s_d\rangle$ to $|g\rangle$, with the latter described by $\sigma_i^-=|g_i\rangle\langle s_{d,i}|$.}
\textcolor{black}{The density matrix $\rho$ therefore obeys the Lindblad master equation}~\cite{Lindblad1976,Gorini1976}
\begin{equation}
{\color{black}
\begin{aligned}
\dot{\rho}=\mathcal{L}[\rho]
=-i[H,\rho]+2\kappa\mathcal{D}[a]\rho
+\gamma\sum_i\mathcal{D}[\sigma_i^-]\rho,
\end{aligned}}
\label{eq:ME}
\end{equation}
\textcolor{black}{Here $\mathcal{L}$ is the Liouvillian operator, $\kappa$ is the cavity decay rate, $\gamma$ is the atomic local dissipation rate, and
$\mathcal{D}[L]\rho\equiv L\rho L^\dagger-\frac{1}{2}\{L^\dagger L,\rho\}$ denotes the Lindblad dissipation operator. Among them, avity decay underlies the open-Dicke superradiant phase structure, while atomic local dissipation breaks spin-length conservation and stabilizes long-time oscillations.}

\paragraph{Mean-field phase diagram.}
Since a full quantum description of the many-body Lindblad dynamics quickly becomes intractable with increasing $N$, we first extract the basic dynamical frame within a mean-field approximation.
Specifically, we neglect correlations and factorize operator products, e.g., $\expval{a\sigma_i^\mu}\simeq\expval{a}\expval{\sigma_i^\mu}$ and $\expval{\sigma_i^\mu\sigma_j^\nu}\simeq\expval{\sigma_i^\mu}\expval{\sigma_j^\nu}$ for $i\neq j$.
\textcolor{black}{The Rydberg interaction enters through the population-dependent effective detuning $\Delta\equiv\omega_a+Vn$, which controls the intrinsic rotation of the transverse spin ($S_x , S_y$) (the complete mean-field equations are given in the Supplemental Material~\cite{SM}).}
\textcolor{black}{At $\gamma=0$, the conservation of the atomic spin length makes the polar fixed points and their instability thresholds analytically tractable.}
\textcolor{black}{In this limit, the two polar fixed points are the $g$-polarized and $s_d$-polarized states, $(\expval{a},\expval{S_z})=(0,\pm N/2)$.}
\textcolor{black}{Their linear instabilities give two critical couplings~\cite{PhysRevA.85.013817,SM}:}
\begin{equation}
\lambda_{\downarrow}^2=\frac{\omega_a(\kappa^2+\omega_c^2)}{4\omega_c},\qquad
\lambda_{\uparrow}^2=\frac{-(\omega_a+V)(\kappa^2+\omega_c^2)}{4\omega_c}.
\label{eq:lambda_c_pm}
\end{equation}
There is a basic asymmetry of our dressed system: the Rydberg-dressed interaction is affected only by the population occupied at  $|s_d\rangle$. Consequently, its Rydberg-dressed shift vanishes at the $g$-polarized pole but is non-zero at the $s_d$-polarized pole, leading to the $V$-independent $\lambda_{\downarrow}$ whereas $\lambda_{\uparrow}$ is renormalized and exists only for $\omega_a+V<0$, which are reflected by Eqs. (\ref{eq:lambda_c_pm}).

We plot the phase diagram of the mean-field \textcolor{black}{dynamics} in parameter space $(V,\lambda)$ in Fig.~\ref{fig:mf}(b).
\textcolor{black}{The two critical couplings divide the phase diagram into a Dicke-like region, a bistable region, and a resonance case.}

(i) \,\emph{Dicke-like region} ($V>-\omega_a$). There is no real-value threshold $\lambda_{\uparrow}$ from Eq. (\ref{eq:lambda_c_pm}) in this regime, so the $s_d$-polarized pole remains unstable. The dynamics therefore follows to the ordinary open Dicke case controlled by $\lambda_{\downarrow}$.

(ii) \,\emph{Bistable region} (\textcolor{black}{$V<-\omega_a$}).
\textcolor{black}{Here $\lambda_{\uparrow}$ is real, so the two polar fixed points can both appear as stable attractors depending on the coupling strength.}
\textcolor{black}{For $-2\omega_a<V<-\omega_a$, $\lambda_{\uparrow}<\lambda_{\downarrow}$, increasing $\lambda$ first destabilizes the $s_d$-polarized pole and then the $g$-polarized pole.}
For $V<-2\omega_a$, the relation between two critical couplings is reversed ($\lambda_{\downarrow}<\lambda_{\uparrow}$), giving a dynamical picture that is opposite to that in the bistable region.
For $\lambda_{\downarrow} < \lambda < \lambda_{\uparrow}$,
the $s_d$-polarized pole is the only stable attractor, so the effective Rydberg interaction provides a convenient route to prepare the $s_d$-polarized state in practice~\cite{Verstraete2009NatPhys,Poyatos1996PRL}.


(iii) \,\emph{Resonance case} ($V=-2\omega_a$). Here the interaction makes the two critical couplings coincide, i.e.,  $\lambda_c\equiv\lambda_{\downarrow}=\lambda_{\uparrow}$, \textcolor{black}{as shown in Fig.~\ref{fig:mf}(b).}
\textcolor{black}{For $\lambda>\lambda_c$, the two polar fixed points lose stability simultaneously, giving a center-like oscillatory dynamics in the mean-field limit.}


\textcolor{black}{We note that, at $\gamma=0$, the conservation of the collective-spin length produces a neutrally stable family of center-like orbits with initial-condition-dependent amplitudes. However, finite local dissipation ($\gamma \neq 0$) can break this conservation and allow trajectories to move between these orbits and relax toward a stable finite-amplitude limit cycle (LC)~\cite{SM}. }

\begin{figure}[tp]
	\centering
	\includegraphics[width=0.98\columnwidth]{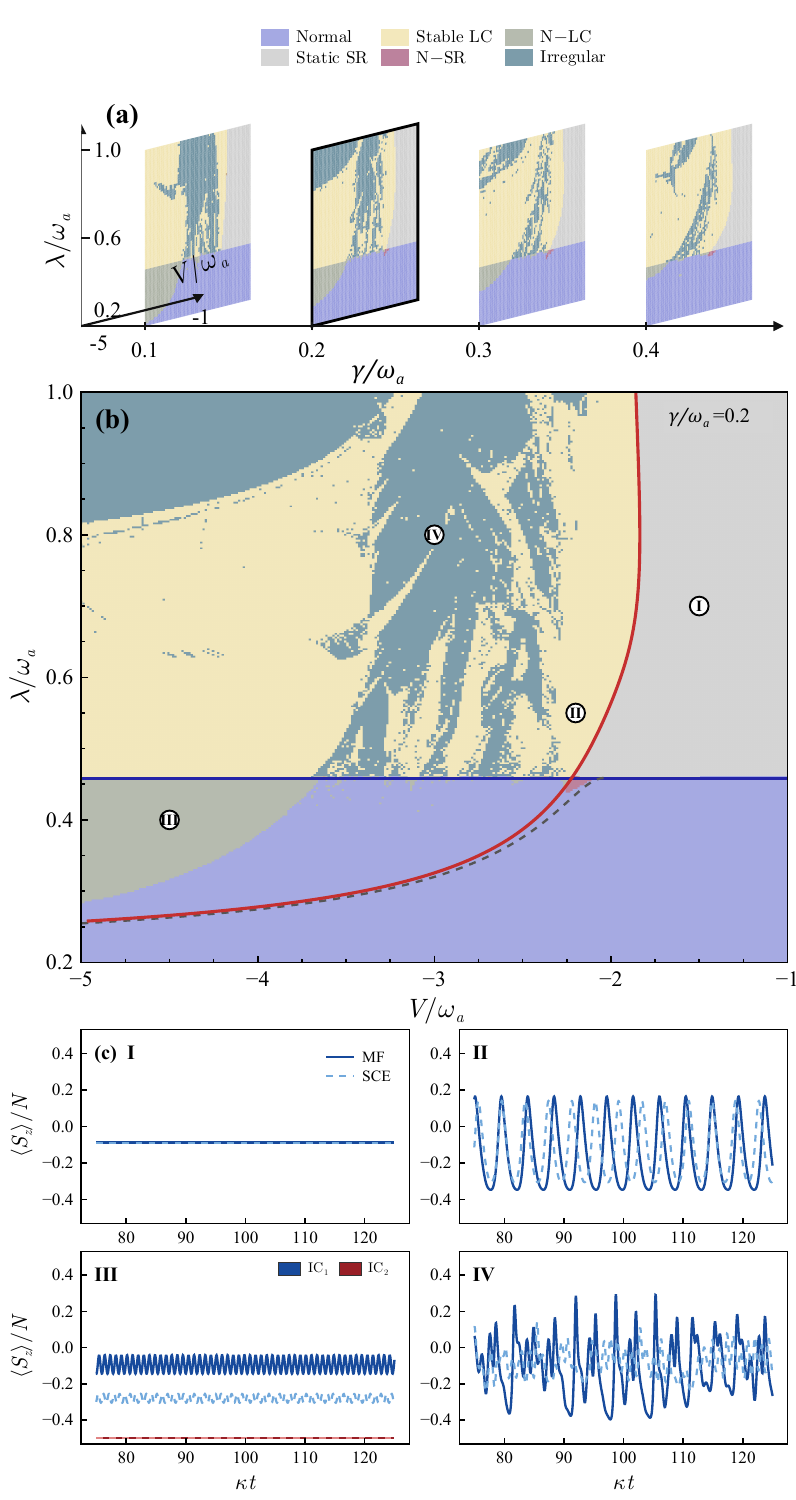}
	\caption{\textcolor{black}{Locally dissipative mean-field phase diagram and representative dynamics.
	(a) Phase diagrams in the $(V,\lambda)$ plane for $\gamma/\omega_a=0.10$, $0.20$, $0.30$, and $0.40$.
	(b) Global phase diagram at $\gamma/\omega_a=0.20$. The grey dashed curves mark the boundary of superradiant-state existence, and the red curves mark the superradiant Hopf boundary, while I--IV denote representative parameter points.
	(c) Mean-field and SCE dynamics in the static superradiant (SR), stable LC, N--LC, and Irregular regimes, respectively.
	The two initial conditions in III resolve the coexisting Normal and finite-amplitude oscillatory attractors.}}
	\label{fig:dissipative_phase}
\end{figure}

\textcolor{black}{As local dissipation breaks conservation of the collective-spin length, the global attractor structure can no longer be inferred from fixed-point instabilities alone. We therefore construct the phase diagrams in Fig.~\ref{fig:dissipative_phase} by evolving the mean-field equations from multiple initial conditions and classifying the resulting long-time attractors. The analytic instability boundaries are overlaid as a local stability skeleton, while the global classification also captures finite-amplitude attractors and basin coexistence that are inaccessible to a purely local analysis.}

\textcolor{black}{We find that the finite local dissipation removes the $s_d$-polarized fixed point and reduces the two-threshold structure to a single normal-state boundary, as shown in Fig.~\ref{fig:dissipative_phase}(a). Because the remaining $g$-polarized fixed point has zero Rydberg population, its critical coupling is independent of $V$ and changes only weakly with $\gamma$. The most pronounced modification occurs in the oscillatory sector at large $|V|$. In the representative slice $\gamma/\omega_a=0.20$ shown in Fig.~\ref{fig:dissipative_phase}(b), a stable LC emerges through the supercritical Hopf instability of the superradiant fixed point. At stronger negative $V$, the Hopf boundary lies below the normal-state threshold $\lambda_c$, producing a coexistence phase of the normal state and a LC (N--LC) below below $\lambda_c$ and leaving the LC as the remaining attractor above it.}

\textcolor{black}{Farther from the Hopf boundary, the dynamics becomes more complex, with stable-LC and irregular regions appearing alternately. We find that the evolution of the population-dependent detuning $\Delta$ plays a central role in organizing the stability of these oscillations. In the mean-field equations, $\Delta$ controls the rotation of the spin. When a trajectory of the system crosses $\Delta=0$, this rotation reverses its direction. Because the cavity field responds on a finite timescale, the reversal changes the relative phase between the spin and cavity field and can destabilize the periodic motion. As $|V|$ is increased further, the detuning range explored by the orbit is shifted entirely below zero, suppressing the sign reversal and allowing the stable LC to recover. This nonlinear spin--cavity feedback captures the re-entrant structure of the phase diagram (a more detailed analysis and Lyapunov analysis is given in the Supplemental Material~\cite{SM}). The representative trajectories in Fig.~\ref{fig:dissipative_phase}(c) illustrate the static-SR, stable-LC, N--LC, and irregular regimes, with the two initial conditions at point III resolving the coexisting Normal and LC attractors. The corresponding SCE trajectories remain qualitatively consistent with the mean-field classification, indicating that these dynamical feature persist upon including leading correlation corrections.}

\paragraph{Continuous time-crystal dynamics.}

\begin{figure}[tp]
	\centering
	\includegraphics[width=0.95\linewidth]{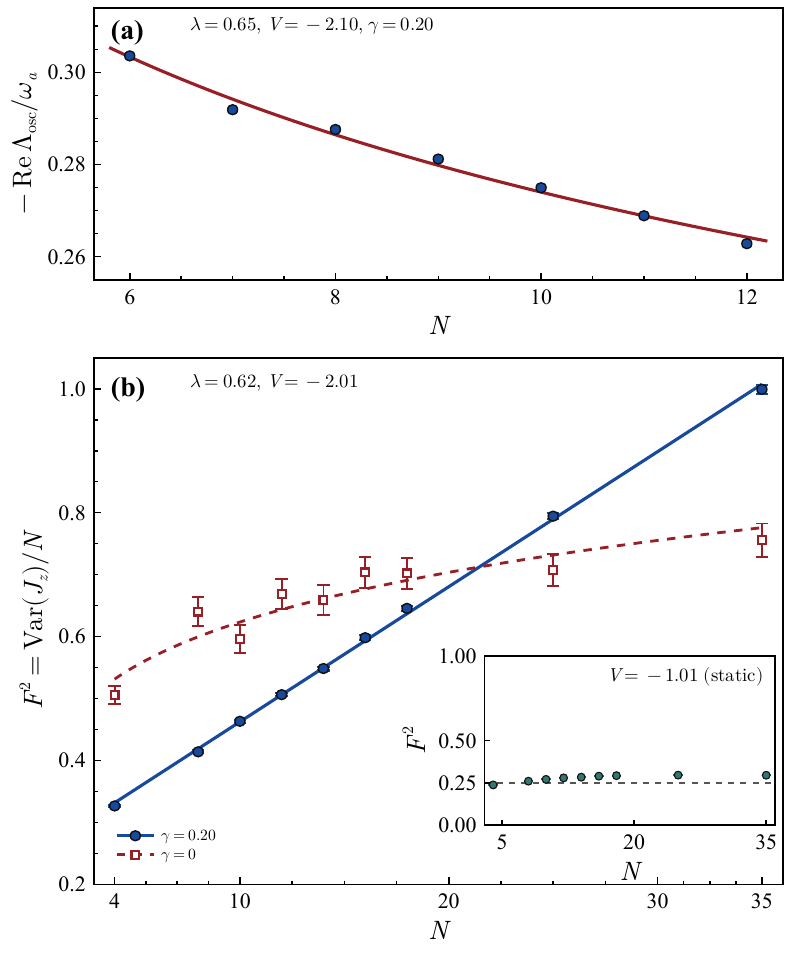}
	\caption{\textcolor{black}{Finite-size quantum signatures of the CTC.
	(a) Damping rate $\Gamma_{\mathrm{osc}}=-\operatorname{Re}\Lambda_{\mathrm{osc}}$ of the continuously tracked oscillatory Liouvillian mode; the solid curve is a power-law fit.
	(b) Connected fluctuations $F^2$ obtained from wave-function Monte Carlo simulations with and without local dissipation at fixed $(\lambda,V)$.
	The solid and dashed curves show linear and power-law fits, respectively. Inset: the static-phase reference.}}
	\label{fig:ctc}
\end{figure}

\textcolor{black}{The persistence of oscillations in the second-order cumulant expansion motivates the examination on whether temporal order extends beyond this finite-order treatment of correlations~\cite{Kubo1962,SM}.}
\textcolor{black}{We therefore analyze the relaxation modes of the Lindblad generator, defined by $\mathcal{L}[R_\mu]=\Lambda_\mu R_\mu$, with each eigenoperator evolving as $e^{\Lambda_\mu t}R_\mu$~\cite{Macieszczak2016PRL}.}
\textcolor{black}{An oscillatory eigenvalue $\Lambda_{\mathrm{osc}}=-\Gamma_{\mathrm{osc}}+i\omega_{\mathrm{osc}}$ thus specifies a damping rate $\Gamma_{\mathrm{osc}}$ and a frequency $|\omega_{\mathrm{osc}}|$.}
\textcolor{black}{Figure~\ref{fig:ctc}(a) shows that the tracked oscillatory mode becomes progressively less damped as $N$ increases, while its frequency remains finite over the accessible sizes~\cite{SM}.}
\textcolor{black}{This increasing lifetime provides finite-size quantum evidence for collective temporal order in the interacting spin-$1/2$ ensemble even in the presence of local dissipation~\cite{Iemini2018PRL,jhd4-1khw}.}

\textcolor{black}{The constructive role of local dissipation is further reflected in the steady-state fluctuations, $F^2=[\langle S_z^2\rangle_{\mathrm{ss}}-\langle S_z\rangle_{\mathrm{ss}}^2]/N$~\cite{Benatti2018JPA}, obtained from wave-function Monte Carlo simulations~\cite{KORNYIK201988,SM}.}
\textcolor{black}{At resonance without including the local dissipation, the sublinear growth of $F^2$ can reflect correlations that increase over accessible sizes, while $\operatorname{Var}(S_z)/N^2$ vanishes under such an extrapolation.}
\textcolor{black}{However, when the local dissipation is considered, $F^2\propto N$, implying $\operatorname{Var}(S_z)\propto N^2$~\cite{RussoPohl2025PRL}.}
\textcolor{black}{One can see that the stable-phase reference approaches $F^2\simeq1/4$, while the trivial independent-spin noise level evolves  with $\operatorname{Var}(S_z)\propto N$ [Fig.~\ref{fig:ctc}(b)].}
\textcolor{black}{The observed scaling features of $F^2$ suggests that local dissipation supports collective fluctuations that persist as the system approaches the thermodynamic limit, paralleling its mean-field role in converting neutral resonant oscillations into an attracting limit cycle.}
\textcolor{black}{Together with the increasingly long-lived finite-frequency mode, this provides complementary finite-size quantum evidence for the CTC.}

\begin{figure}[tp]
	\centering
	\includegraphics[width=\linewidth]{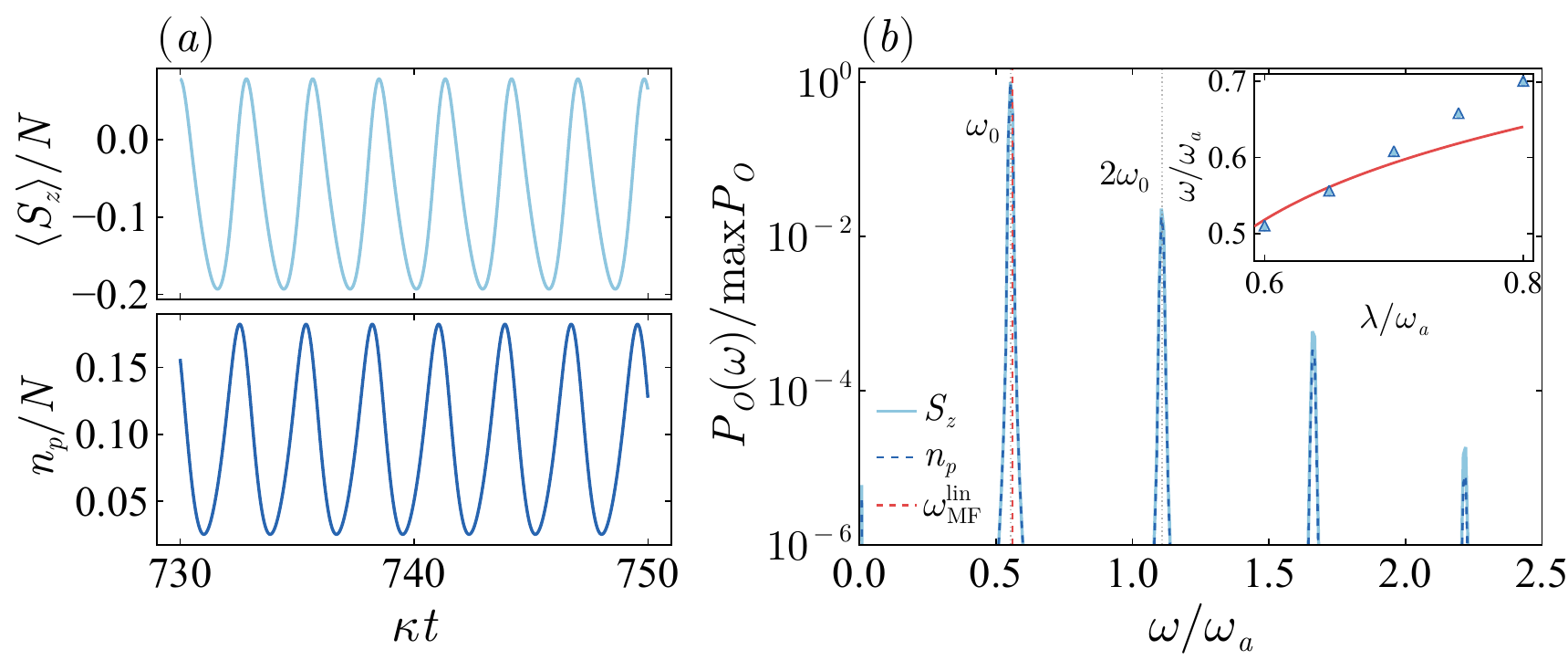}
	\caption{\textcolor{black}{Optical readout of collective oscillations under local dissipation.
	(a) Late-time SCE dynamics of the normalized atomic inversion and intracavity photon number.
	(b) Corresponding mean-subtracted power spectra, each normalized to its maximum at nonzero frequency.
	Gray dotted lines mark $\omega_0$ and $2\omega_0$, and the red dashed line marks $\omega_{\mathrm{MF}}^{\mathrm{lin}}$.
	Inset: $\omega_{S_z}$ and $\omega_{n_p}$ versus $\lambda$, together with the tracked mean-field linearization branch.
	}}
	\label{fig:ctcfreq}
\end{figure}

\textcolor{black}{Beyond these finite-size signatures, the quantized cavity field provides optical access to the collective atomic dynamics.}
\textcolor{black}{We denote by $\omega_{S_z}$ and $\omega_{n_p}$ the dominant frequencies extracted from the late-time SCE dynamics of $\expval{S_z(t)}$ and $n_p(t)\equiv\expval{a^\dagger a}$, respectively.}
\textcolor{black}{In the locally dissipative regime shown in Fig.~\ref{fig:ctcfreq}, these frequencies coincide within numerical resolution, $\omega_{S_z}\simeq\omega_{n_p}$, showing that the collective atomic oscillations are imprinted on the intracavity photon number.}
\textcolor{black}{As an independent reference, we define $\omega_{\mathrm{MF}}^{\mathrm{lin}}=|\operatorname{Im}\mu_{\mathrm{osc}}|$, where $\mu_{\mathrm{osc}}$ is the oscillatory eigenvalue obtained by linearizing the dissipative mean-field equations about the corresponding superradiant fixed point~\cite{SM}.}
\textcolor{black}{This small-amplitude frequency captures the SCE result near the Hopf boundary, while deviations at stronger coupling mark the limits of the linearized description.}
\textcolor{black}{The cavity output thus provides an experimentally accessible readout of the collective temporal dynamics.}

\paragraph{Conclusion.}

\textcolor{black}{We have shown that a Rydberg-interacting open Dicke model with local dissipation supports a rich dissipative phase structure, including stable LC, N--LC coexistence, and irregular oscillatory regimes.}
\textcolor{black}{Local dissipation commonly damps collective oscillations in spin-$1/2$ systems, while recent constructive approaches have employed spin-$1$ or other three-level structures with competing excitation branches~\cite{jhd4-1khw,RussoPohl2025PRL}.}
\textcolor{black}{Our results demonstrate a distinct route in which local dissipation provides amplitude selection and acts jointly with the population-dependent Rydberg detuning and feedback from a lossy quantized cavity.}
\textcolor{black}{The decreasing damping of a finite-frequency Liouvillian mode and the near-linear connected-fluctuation scaling provide complementary finite-size quantum evidence for the resulting temporal order.}
\textcolor{black}{Consequently, our framework provides an unique cavity-assisted interacting spin-$1/2$ platform for continuous time-crystalline dynamics.}

\textcolor{black}{The tunability of the Rydberg interaction, together with direct optical access through the cavity output, makes this platform well suited for probing nonequilibrium collective dynamics.}
\textcolor{black}{In particular, the coexistence of stationary and oscillatory attractors provides a setting for studying dynamical criticality, multistability, and initial-state selection among distinct long-time states~\cite{Ding2022NatPhysMetrology,Liu2025NatCommun,Wu2024NatPhys}.}
\textcolor{black}{Extending the system beyond the all-to-all approximation to spatially structured Rydberg interactions would reveal how finite interaction ranges and spatial correlations compete with collective cavity feedback~\cite{PhysRevLett.134.213604,Weckesser2025Science,BrowaeysLahaye2020NatPhys}.}
\textcolor{black}{Introducing multiple cavity modes would add further collective channels and spatial mode structure, opening a route to more complex dynamical phases and spatiotemporal order~\cite{PRXQuantum.3.020348,PhysRevLett.133.106901,Tay2025NatCommun}.}

\begin{acknowledgments}
This work is supported by the National Key Research and Development Program of China (no. 2023YFA1407200) and the National Natural Science Foundation of China (12192252).

\paragraph{Data availability}
The data that support the findings of this article are not publicly available. The data are available from the authors upon reasonable request.

\end{acknowledgments}

\bibliographystyle{apsrev4-2}
\bibliography{references}

\end{document}